\magnification\magstep2

\font\logo=logo10 
\def\MF{{\logo META}\-{\logo FONT}}
\def\TeX{T\hbox{\hskip-.1667em\lower.424ex\hbox{E}\hskip-.125em X}}
\def\display#1:#2:#3\par{\par\hangindent #1 \noindent
	\hbox to #1{\hfill #2 \hskip .1em}\ignorespaces#3\par}

\line{\bf THEORY and PRACTICE\hfill}
\bigskip
\line{Keynote address for the 11th World Computer Congress\hfill}
\line{(Information Processing 89)\hfill}
\line{San Francisco, 28 August 1989\hfill}
\bigskip
\line{[SLIDE 0 to be shown during introduction of the speaker]\hfill}
\medskip

\noindent
Good morning! I want to welcome you all to the San Francisco Bay area
and to nearby Silicon Valley, where I~live at Stanford University.
[SLIDE~1] In recent years the people around here have been taking
advantage of an idea that originated, I~think, in the international
road signs that have spread from Europe to the rest of the world:  the
idea of {\it icons}, as graphic representations of information. Icons
have now become so pervasive, in fact, that I~think people might soon
be calling this place Silly Icon Valley!

The title of my talk this morning is {\it Theory and Practice}, and in
order to be up-to-date I want to begin by showing you two icons that
might make suitable pictographs for the notions of theory and
practice.
I~didn't have any trouble finding such images, because the reference
section of our local telephone directory contains lots of icons these
days. Looking at those pages, I~immediately spotted an image that
seems just right to depict theory: [SLIDE~2]\ \ A~light bulb of
inspiration. And what about practice? Right next to that light bulb in
the phone book was another suitable image: [SLIDE $2+3$]\ \ A~hand
carrying a briefcase.

Theory and Practice. Both of these English words come from the Greek
language, and their root meanings are instructive. [SLIDE~4] The Greek
$\theta\epsilon\omega\rho\acute\iota\alpha$ means seeing or viewing, while
$\pi\rho\alpha\kappa\tau\iota\kappa\acute\eta$ means doing, performing. We
owe to ancient Hellenic philosophers the revolutionary notion of
theory as the construction of ideal mental models that transcend
concrete physical models. They taught us systems of logic by which
intuitive assumptions and rules of inference can be made explicit;
therefore significant statements can be rigorously and conclusively
proved. Throughout the ages, practitioners have taken such theories
and applied them to virtually every aspect of civilization. [SLIDE
$4+5$]
\ \  Thus, we can say that theory is to practice as rigor is to vigor.

Theory {\it and\/} Practice. [SLIDE $6+7$]\ \ The English word `and'
 has several meanings, one of which corresponds to the mathematical
notion of `plus'. When many people talk about theory and practice,
they are thinking about the sum of two disjoint things. In a similar
way, when we refer to `apples and oranges', we're talking about two
separate kinds of fruit.

[SLIDE $6+8$]\ \ But I wish to use a stronger meaning of the word
`and', namely the logician's notion of `both and', which corresponds
to the {\it intersection\/} of sets rather than a sum. The main point
I want to emphasize this morning is that both theory and practice can
and should be present simultaneously. Theory and practice are not
mutually exclusive; they are intimately connected. [SLIDE $6+8+9$]\ \ 
They live together and support each other.

This has always been the main credo of my professional life. I~have
always tried to develop theories that shed light on the practical
things I~do, and I've always tried to do a variety of practical things
so that I~have a better chance of discovering rich and interesting
theories. It seems to me that my chosen field, computer
science---information processing---is a field where theory and
practice come together more than in any other discipline, because of
the nature of computing machines.

I came into computer science from mathematics, so you can suspect that
I have a soft spot in my heart for abstract theory. I~still like to
think of myself as a mathematician, at least in part; but during the
1960s I~became disenchanted with the way mathematics was going. I'll
try to explain why, by saying a few things about the history of
mathematical literature. [SLIDE~10]\ \ The first international journal
of mathematics was founded in 1826 by a man named August Leopold
Crelle. I~think its title was significant: ``{\sl Journal f\"ur die
reine und angewandte Mathematik\/}', a~journal for pure and applied
mathematics. In many people's eyes, `pure mathematics' corresponds to
`theory' and `applied  mathematics' corresponds to `practice'; so
there we have it, theory and practice. This venerable journal is
still being published today, currently in volume number~398.
[SLIDE~11]\ \ Another journal with the equivalent title in French
began publication ten years later. This one too has continued to the
present day, and both journals still mention both pure and applied
mathematics in their titles. But there was a time when the only
applied mathematics you could find in these journals consisted of
applications to pure mathematics itself!

[NO SLIDE]\ \ When theory becomes inbred---when it has grown several
generations away from its roots, until it has completely lost touch
with the read world---it degenerates and becomes sterile. I~was
attracted to computer science because its theory seemed much more
exciting and interesting to me than the new mathematical theories I
was hearing about in the~60s. I~noticed that computer science theory
not only had a beautiful abstract structure, it also answered
questions that were relevant to things I wanted to do. So I became a
computer scientist.

History teaches us that the greatest mathematicians of the past
combined theory and practice in their own careers. For example, let's
consider Karl Friedrich Gauss, who is often called the greatest
mathematician of all time, based on the deep theories he discovered.
[SLIDE~12]\ \ Here is an excerpt from one of his diaries; Gauss left
behind thousands of pages of detailed computations. His practical work
with all these numbers led him to discover the method of least squares
and the so-called Gaussian distribution of numerical errors.
[SLIDE~13]\ \ He also made measurements of the earth and drew this map
as a basis for land surveys in parts of Germany, the Netherlands, and
Denmark. [SLIDE~14]\ \ His study of magnetism led him to publish a
series of world maps such as this one. Thus Gauss was by no means
purely a theoretician. His practical work went hand in hand with his
theoretical discoveries in geometry and physics.

One of the main reasons I've chosen to speak about Theory and Practice
this morning is that I've spent the past 12~years working on a project
that 
has given me an unusual opportunity to observe how theory and practice
support each other. [NO SLIDE]\ \ My project at Stanford University has
led to the development of two pieces of software called \TeX\ and 
\MF\kern-1pt: \TeX, a system for typesetting, and \MF\kern-1pt,
a~system for generating alphabets and symbols. [SLIDE~15]\ \ Here are
the icons for \TeX\ and \MF\kern-1pt.

Throughout my experiences with the \TeX\ project, I couldn't help
noticing how important it was to have theory and practice present
simultaneously in equal degrees. One example of this is the method for
hyphenating words that was discovered by my student Frank Liang.
[SLIDE~16]\ \ Suppose we want to find permissible places to break up
the word `hyphenation'. Liang's idea is to represent hyphenation rules
by a set of patterns, where each pattern is a string of letters
separated by numerical values. We find all the patterns that appear as
substrings of the given word, as shown here; and then we calculate the
maximum of all the numbers that occur between each pair of adjacent
letters. If the resulting number is odd, it represents a place to
break the word; but if it is even, we don't insert a potential hyphen.

The beauty of Liang's method is that it is highly accurate, it runs
fast, and it takes up very little space inside a computer. Moreover,
it works with all languages, not just English: Successful sets of
patterns have already been found for French, German, Italian, Spanish,
Portuguese, Swedish, Icelandic, Russian, and other languages. Thus, it
is a uniform method able to support international communication. Liang
discovered this unified method only after considerable theoretical
study of other techniques, which solved only special cases of the
problem. And his practical work also had a theoretical payoff, because
it led him to discover a new kind of abstract data structure called a
dynamic trie, which has turned out to be of importance in other
investigations. I~think it's reasonable to compare this with some of
Gauss's work; Gauss worked with masses of numerical data while Liang
worked with masses of linguistic data, but in both cases there was an
enrichment of practice that would have been impossible without the
theory and an enrichment of theory that would have been impossible
without the practice.

[NO SLIDE]\ \ That was an example from \TeX; let me give another
example, this time from \MF\kern-1pt. One of the key problems of
discrete geometry is to draw a line or curve that has approximately
uniform thickness although it consists entirely of square pixels. The
obvious way to solve this problem is to draw a solid line of the
desired thickness, without thinking about the underlying raster, and
then to digitize the two edges of that line separately and fill in the
region inside. But this obvious approach doesn't work. [SLIDE~17]\ \
For example, here are two straight lines of slope 1/2 and thickness~1
that were drawn by the obvious method. When we digitize the two edges
and fill the inner region, [SLIDE $17+18$]\ \  the lower line comes
out 50\% darker than the upper one, because it happens to fall in a
different place on the raster.

There's a better way, which I'll call the diamond method. Imagine a
diamond-shaped pen tip, one pixel tall. [SLIDE~19]\ \ Draw a line or
curve with this pen, and then digitize the edges. Now you get a line
or curve that has nearly uniform thickness, regardless of where it
falls on the raster. The ``obvious'' method I mentioned before
corresponds to lines that you would draw when the tip of the pen is a
circle of diameter~1 instead of a diamond. The track of a circular pen
nib does not digitize well, but the track of a diamond-shaped pen nib
does.

[SLIDE~20]\ \  Here's another example, using circular and
diamond-shaped pens to draw a circle whose radius is slightly greater
than~7.5. In this case the circular pen gives a digital track [SLIDE
$20+21$]\  that's noticeably heavier when it travels diagonally than
when it is travelling horizontally or vertically. The diamond pen
gives a much nicer digital circle without such glitches.

My student John Hobby found a beautiful way to extend  these ideas  to
curves of greater thickness. [SLIDE~22]\ \ Here, for example, is an
octagon-shaped pen nib that turns out to give the best results when
you want to draw curves that are slightly less than 3~pixels thick.
Hobby developed \MF's polygonal method of curve drawing by creating a
truly elegant combination of number theory and geometry. His work
is one of the nicest blends of theory and practice I~have ever seen:
It's a case where deep theoretical results have made an important
contribution to a practical problem, and where the theory could only
have been discovered by a person who was thoroughly familiar with both
the practice of digitization and with mathematical theories that had
been developed for quite different problems.

[NO SLIDE]\ \ I want to mention also a third example, This one isn't as
important as the other two, but I can't resist telling you about it
because I just thought of it four days ago. I~decided last week to
make some extensions to \TeX\ so that it will be more useful for
languages other than English. The new standard version of \TeX\ will
support 8-bit character sets instead of only the 7-bit ASCII code.
Furthermore it will allow you to hyphenate words from several
different languages within the same paragraph, using different sets of
patterns for each language. One of the new features will be an
extension of the mechanism by which \TeX\ makes ligatures in the text,
and that's the method I want to explain now.

Suppose two letters occur next to each other in a word that is to be
typeset by the computer; I'll call those letters $\alpha$
and~$\omega$.
[SLIDE~23]\ \ The present version of \TeX\ allows the font designer to
say that the letters $\alpha$ and~$\omega$ should be replaced by a
ligature, say~$\lambda$. This is the way, for example, that an~`f'
followed by an~`i' is converted into a symbol for `fi' that looks
better.

The new version of \TeX\ will extend this mechanism as follows. A~new
letter~$\lambda$ will be inserted between $\alpha$ and~$\omega$, and
the original letters might still remain. [SLIDE~24]\ \  There are nine
cases, depending on what letters are kept and depending on where \TeX\
is instructed to look next for another possible ligature. (The little
caret between letters in this picture shows where \TeX\ is focussing
its attention.) The first case here shows the old ligature mechanism;
the middle seven cases are new; and the bottom case is the normal
situation where no ligature is to be inserted.

This new mechanism has a potential danger. A careless user can now
construct ligature instructions that will get \TeX\ into an infinite
loop. [SLIDE~25]\ \ For example, suppose we have the four rules shown
at the top of this illustration. Then when `a' is followed by~`z', the
rules set off a chain reaction that never stops.

To minimize this danger, I need an algorithm that will take a given
set of ligature rules and decide if it can spawn an infinite loop. And
that's where computer science theory comes to the rescue! [SLIDE~26]\
\ We can define a function~$f$ on letter pairs according to the nine
ligature possibilities, as shown here. This definition is recursive.
It's not hard to see that $f$ is well defined if and only if there are
no infinite ligature loops; we can understand this from the theory of
deterministic pushdown automata. (The value of $f(\alpha,\omega)$
represents the letter just preceding the cursor when the  cursor first
moves to the right of the original~$\omega$.) And we can check whether
or not $f$ is well defined by using a small extension of an important
algorithm called depth-first search.

I like this example not only because it gives an efficient,
linear-time algorithm for testing whether or not a ligature loop
exists. This practical problem also showed me how to extend the theory
of depth-first search in a way that I hadn't suspected before. And I
have a hunch the extended theory will have further ramifications,
probably leading to additional applications having nothing to do with
ligatures or typesetting.

What were the lessons I learned from so many years of intensive work
on the practical problem of setting type by computer? One of the most
important lessons, perhaps, is the fact that SOFTWARE IS HARD.
[SLIDE~27]\ \ From now on I shall have significantly greater respect
for every successful software tool that I encounter. During the past
decade I was surprised to learn that the writing of programs for \TeX\
and for \MF\ proved to be much more difficult than all the other
things I had done (like proving theorems or writing books). The
creation of good software demands a significantly higher standard of
accuracy than those other things do, and it requires a longer
attention span than other intellectual tasks.

My experiences also strongly confirmed my previous opinion that THE
BEST THEORY IS INSPIRED BY PRACTICE and THE BEST PRACTICE IS INSPIRED
BY THEORY. [SLIDE~28]\ \ The examples I've mentioned, and many others,
convinced me that neither theory nor practice is healthy without the
other.

But I don't want to give the impression that theory and practice are
just two sides of the same coin. No. They deserve to be mixed and
blended, but sometimes they also need to be pure. I've spent many an
hour looking at purely theoretical questions that go way beyond any
practical application known to me other than sheer intellectual
pleasure. And I've spent many an hour on purely practical things like
pulling weeds in the garden or correcting typographic errors, not
expecting those activities to improve my ability to discover
significant theories. [SLIDE~29]\ \ Still, I believe that most of the
purely practical tasks I undertake do provide important nourishment
and direction for my theoretical work; and I believe that the hours I
spend contemplating the most abstract questions of pure mathematics do
have a payoff in sharpening my ability to solve practical problems.

When I looked for an icon that would be appropriate for `practice',
I~was tempted to use another one instead of the briefcase---a symbol
for money! [SLIDE $29+30$]\ \ It seems that people who do practical
things are paid a lot more than the people who contribute the
underlying theory. Somehow that isn't right. The past decade has, in
fact, witnessed a very unfortunate trend in the patterns of funding
for basic, theoretical research. We used to have a pretty well
balanced situation in which both theory and practice were given their
fair share of financial support by enlightened administrators. But in
recent years, greater and greater amounts of research dollars have been
switched away from basic research and earmarked for mission-oriented
projects. The people who set the budgets have lost consciousness of
the fact that the vast majority of the crucial ideas that go into the
solution of these mission-oriented problems were originally discovered
by pure scientists, who were working alone, independently, on basic
research. At the present time the scientific community faces a crisis
in which a substantial number of the world's best scientists in all
fields cannot get financial support for their work unless they
subscribe to somebody else's agenda telling them what to do. We need
to go back to a system where people who have demonstrated an ability
to devise significant new theories are given a chance to set their own
priorities. We need a lot of small projects devised by many
independent scientists, instead of concentrating most of our resources
on a few huge projects with predefined goals. [SLIDE $29+30+31$]\ \ 
In other words, we need a balance between theory and practice in the
budgets for scientific research, as well as in the lives of individual
scientists. Otherwise we'll face a big slump in our future abilities
to tackle new problems.

These comments hold true for industry as well as for the university
community. Many of the graduates of Stanford's Computer Science
Department who have written Ph.D. theses about theoretical subjects
have now taken jobs in Silicon Valley and elsewhere; and they have in
most cases been able to work with enlightened managers who encourage
them to continue doing basic research. I~think it's fair to state that
these so-called theoreticians are now considered to be among the key
employees of the companies for which they work.

[SLIDE 32]\ \  Speaking of key employees reminds me that this is a
keynote speech; indeed, this morning is surely the only time in my
life when I'll be able to give the keynote address to an IFIP
Congress. 
So I would like to say something memorable, something of value,
something that you might not have expected to hear. I~thought about
David Hilbert's famous address to the International Congress of
Mathematicians in 1900, when he presented a series of problems as
challenges for mathematicians of the 20th century. My own goals are
much more modest than that; but I {\it would\/} like to challenge some
of you in the audience to combine theory and practice in a way that I
think will have a high payoff.

[SLIDE 33]\ \  My challenge problem is simply this: {\it Make a
thorough analysis of everything your computer does during one second
of computation}. The computer will execute several hundred thousand
instructions during that second; I'd like you to study them all. The
time when you conduct this experiment should be chosen randomly; for
example, you might program the computer itself to use a random number
generator to decide just what second should be captured and recorded.

Many people won't be able to do this experiment easily, because they
won't have hardware capable of monitoring its own activities. But I
think it should be possible to design some tracing software that can
emulate what the machine would have done for one second if it had been
running freely.

Even when the machine's instructions are known, there will be
problems.
The sequence of operations will be too difficult to decipher unless
you have access to the source  code from which the instructions were
compiled. University researchers who wish to carry out such an
experiment would probably have to sign nondisclosure agreements in
order to get a look at the relevant source code. But I want to urge
everyone who has the resources to make such a case study to do so, and
to compare notes with each other afterward, because I am sure the
results will be extremely interesting; they will tell us a lot about
how we can improve our present use of computers.

I discussed this challenge problem with one of the botanists at
Stanford, since I know that biologists often make similar studies of
plant and animal life in a randomly chosen region. [SLIDE~34]\ \ She
referred me to a recent project done in the hills overlooking
Stanford's campus, in which all plants were identified in several
square blocks of soil. The researchers added fertilizer to some of the
plots, in an attempt to see what this did to the plant life. Sure
enough, the fertilizer had a significant effect on the distribution of
species. 

[SLIDE 35]\ \ My colleague also told me about another recent
experiment in which British researchers identified and counted each
tree in a tropical rain forest. About 250,000 trees were enumerated
altogether. I~imagine a typical computer will execute something like
that number of instructions every second, so my specification of
exactly one computer second seems to be reasonable in scale.

Here are some of the questions I would like to ask about randomly
captured seconds of computation: [SLIDE 36]

\smallskip
\display 30pt:$\bullet$:
Are the programs correct or erroneous? (I have to report reluctantly
that nearly every program I have examined closely during the past
thirty years has contained at least one bug.)

\smallskip
\display 30pt:$\bullet$:
Do the programs make use of any nontrivial theoretical results?

\smallskip
\display 30pt:$\bullet$:
Would the programs be substantially better if they made more use of
known theory? Here I am thinking about theories of compiler
optimization as well as theories of data structures, algorithms,
protocols, distributed computation, and so on.

\smallskip
\display 30pt:$\bullet$:
Can you devise new theoretical results that would significantly
improve the performance of the programs during the second in question?

\smallskip
\noindent
In a sense, I'm asking questions something like the botanists
considered: I'm asking to what extent computer programs have been
``fertilized'' by theory, and to what extent such fertilization and
cross-pollination might be expected to improve our present situation.
I~hope many of you will be inspired to look into questions like this.

[SLIDE 37]\ \ In conclusion, let me encourage all of you to strive for
a healthy balance between theory and practice in your own lives. If
you find that you're spending almost all your time on theory, start
turning some attention to practical things; it will improve your
theories. If you find that you're spending almost all your time on
practice, start turning some attention to theoretical things; it will
improve your practice.

The theme of this year's IFIP Congress is Better Tools for
Professionals. I~believe that the best way to improve our tools is to
improve the ways we blend Theory with Practice. Thank you for
listening.

\bigskip
\noindent
References

\bigskip
\display 30pt:$\bullet$:
Pacific Bell SMART Yellow Pages for Palo Alto, Redwood City and Menlo
Prk, May 1989--90, page A47.

\smallskip
\display 30pt:$\bullet$:
K. F. Gauss's {\sl Werke}, volumes $10^1$ and 12.

\smallskip
\display 30pt:$\bullet$:
Franklin Mark Liang, ``Word hy-phen-a-tion by com-pu-ter,'' Ph.D.
dissertation, Stanford University, September 1985.

\smallskip
\display 30pt:$\bullet$:
John Douglas Hobby, ``Digitized brush trajectories,'' Ph.D.
dissertation, Stanford University, September 1985.

\smallskip
\display 30pt:$\bullet$:
Donald E. Knuth, {\sl Computers \& Typesetting}, five volumes,
Addison\kern.1em--Wesley, 1986.

\smallskip
\display 30pt:$\bullet$:
Richard J. Hobbs, S. L. Gulmon, V. J. Hobbs, and H. A. Mooney,
``Effects of fertilizer addition and subsequent gopher disturbance on
a serpentine annual grassland community,'' {\sl{\OE}cologia\/ \bf 75}
(1988), 291--295.

\smallskip
\display 30pt:$\bullet$:
Stephen P. Hubbell and Robin B. Foster, ``Canopy gaps and the dynamics
of a neotropical forest,'' in {\sl Plant Ecology}, ed.\ by Michael~J.
Crawley, Blackwell Scientific, 1986.

\bigskip
\noindent
The preparation of this report was supported in part by National
Science Foundation grant CCR--8610181. `\TeX' is a trademark of the
American Mathematical Society. `\MF' is a trademark of
Addison\kern.1em--Wesley Publishing Company.

\bye